# An Inside Look at the Ti-MoS$_2$ Contact in Ultra-thin Field Effect Transistor with Atomic Resolution

*Ryan J. Wu[1+], Sagar Udyavara[1+], Rui Ma[2], Yan Wang[3], Manish Chhowalla[3], Steven J. Koester[2], Matthew Neurock[1]\*, K. Andre Mkhoyan[1]\**

[1]Department of Chemical Engineering and Material Science, University of Minnesota, Minneapolis, MN 55455

[2]Department of Electrical and Computer Engineering, University of Minnesota, Minneapolis, MN 55455

[3]Department of Material Science and Engineering, Rutgers University, Piscataway, NJ 08854

[+] These authors contributed equally to this work

*Corresponding authors. E-mail: mneurock@umn.edu (MN) and mkhoyan@umn.edu (KAM)

**Abstract:** Two-dimensional molybdenum disulfide (MoS$_2$) is an excellent channel material for ultra-thin field effect transistors. However, high contact resistance across the metal-MoS$_2$ interface continues to limit its widespread realization. Here, using atomic-resolution analytical scanning transmission electron microscopy (STEM) together with first principle calculations, we show that this contact problem is a fundamental limitation from the bonding and interactions at the metal-MoS$_2$ interface that cannot be solved by improved deposition engineering. STEM analysis in conjunction with theory shows that when MoS$_2$ is in contact with Ti, a metal with a high affinity to form strong bonds with sulfur, there is a release of S from Mo along with the formation of small Ti/Ti$_x$S$_y$ clusters. A destruction of the MoS$_2$ layers and penetration of metal can also be expected. The design of true high-mobility metal-MoS$_2$ contacts will require the optimal selection of the metal or alloy based on their bonding interactions with the MoS$_2$ surface. This can be advanced by evaluation of binding energies with increasing the number of atoms within metal clusters.

Transition metal dichalcogenides (TMDs), with the chemical formula $MX_2$, where M is a group IV, V, or VI transition metal and X is the chalcogenide, make up a sub-group of materials with highly tunable electronic properties within the family of van der Waals (vdW) bonded two-dimensional (2D) materials. Ultra-thin field effect transistors (FETs) using molybdenum disulfide ($MoS_2$) as the channel material have shown excellent performance, making them viable for sub-10 nm node devices (*1-5*). However, optimizing the contact between the 2D semiconductor and the 3D metal electrode with direct chemical bonding remains a challenge due to the high contact resistance (*6-9*) commonly attributed to Fermi level pinning at the metal-$MoS_2$ interface (*10-12*). Experimental and computational reports have shown or suggested that surface states created from adsorbed contaminants and damage by kinetic energy transfer from metal deposition can pin the Fermi level and increase the contact resistance (*13, 14*). Therefore, it is largely believed that continued improvement of the metal deposition and FET fabrication will lead to "clean" metal-TMD interfaces without pinning (*11, 15, 16*). However, whether this challenge is one in optimizing the deposition of the metal on $MoS_2$ or, instead, due to the fundamental bonding and chemistry at metal-$MoS_2$ interface is still unknown. While it is possible to engineer metal-$MoS_2$ contact with vdW interactions either by incorporating a intermediate layer (*17, 18*), by transferring an entire metal film (*5*), or by using a metal with very low affinity to bond with sulfur, the transmittance of charge carriers will be limited by the vdW barrier. Ultimately, moderate interatomic bonding that anchors the metal to the $MoS_2$ without restructuring the metal-$MoS_2$ interface is needed for true high-mobility charge transfer in $MoS_2$ embedded FETs.

Here, using atomic-resolution analytical scanning transmission electron microscopy (STEM), we directly image and characterize a contamination-free, deposition damage-free interface between Ti metal electrode and an $MoS_2$ channel in a FET to address the issue of pinning. Ti possesses a very strong affinity for sulfur with a Ti-S bond dissociation energy of -3.69 eV (vs. that of Mo-S at -2.35 eV) (*19*). Consequentially, it is often used as the metal electrode as it provides a strong contact with the $MoS_2$ surface which is desirable for effective charge injection into the $MoS_2$ (*20-22*). Annular dark-field (ADF)-STEM imaging is used in conjunction with electron energy-loss spectroscopy (EELS) to measure the changes in the atomic structure as well as the electronic structure at each layer of MoS$_2$. First principle density functional theory (DFT) calculations are carried out to understand these observed changes in atomic and electronic structure of the Ti-MoS$_2$ contact and to help identify possible paths to improve metal-MoS$_2$ interface performance.

Figure 1(a) shows the schematic layout of the FETs studied here, where the Ti layer from the metal contact directly interfaces with the MoS$_2$ channel. A low-magnification cross-sectional ADF-STEM image of such a FET is shown in Figure 1(b) where the Au/Ti contact and Si/SiO$_2$ substrate bookend the layers of the device. Only some of the individual layers are visible at this magnification. A higher-magnification image of the FET, where the metal contact, MoS$_2$ channel, and all other structural layers are visible, is provided in the SI (see Figure S1). EELS spectra of the oxygen K edge measured from the Ti-MoS$_2$ interfacial layers show a negligible oxygen content (see Figure S2 in SI), which rules out possible water or oxide contaminants (*15*) at the Ti-MoS$_2$ interface during Ti deposition. Under the low-energy conditions of electron beam evaporation-dependent physical vapor deposition, as discussed in the Methods section of the SI, no physical damage to MoS$_2$ layers should be occurring due to the Ti deposition.

Figure 1(c) shows a high-magnification ADF-STEM image of the Ti-MoS$_2$ interface, where the individual MoS$_2$ layers can be clearly identified. As can be seen at this magnification, the Ti layer forms a non-uniform interface with the MoS$_2$. Along the Ti-MoS$_2$ interface, there are clear identifiable areas where the Ti extends into the MoS$_2$ channel and visibly alters the structure of the topmost layer. Next to them are areas where the Ti retreats and small void pockets are formed, leaving the topmost MoS$_2$ layer pristine. This non-uniformity suggest that the Ti atoms tend to cluster on MoS$_2$ during deposition, a behavior suggested or predicted before (*16*) but never directly observed experimentally (see Figure S3 in SI for additional images).

The atomic-resolution ADF-STEM image in Figure 1(d) obtained from one of these Ti-clustered areas shows that the topmost MoS$_2$ layer is indeed completely degraded and barely identifiable. It also shows that the intensity of the ADF signal in the Ti region directly above the 1$^{st}$ MoS$_2$ layer is lower and gradually increases with distance away from the MoS$_2$. This suggests that the Ti is less dense directly above the MoS$_2$, an observation with two possible explanations: (i) the presence of more void pockets along the electron beam direction, which is likely, or (ii) the presence of a Ti sublayer with lower atomic density. The viability of the second possibility will be considered later. These observations combined with earlier predictions suggest that the non-uniformity of the Ti-MoS$_2$ interface may be the result of chemical bonding between Ti and S that drives a restructuring of the Ti-MoS$_2$ interface to form more energetically-favored structures and, as such, would be intrinsic to the Ti contact.

To measure the changes to the electronic structure of MoS$_2$ caused by the Ti contact, layer-by-layer EELS analysis was performed across the Ti-MoS$_2$ interface. Electronic excitations from the sulfur core 2$p$ states to the empty states of conduction band are recorded from each MoS$_2$ layer using core-level EELS. These experiments measure changes in the local electronic density of states (DOS) of the conduction band, which are sensitive to the local bonding environment of the probed atoms (*23*). An example of such a core-level EELS dataset is presented in the SI (Figure S4). Figure 2(a) shows two S L$_{23}$ core-level EELS edges measured from the MoS$_2$ channel: one from layer 1, the topmost MoS$_2$ layer directly in contact with Ti, and the other from layer 5, which is far away from the interface and represents the bulk-like MoS$_2$ layer. The dominating features of the S L$_{23}$ edge fine structure, peaks I and II, composed of S 3$s$ and 3$d$ partial-DOS (*24*) are quantifiably different in these measured spectra, as can be seen in the difference spectrum shown in Figure 2(a). In comparison, these peaks are more subdued in layer 1 - a strong indication that the atomic structure of this layer has lost its periodicity considerably (*23*), which is consistent with observations from the ADF-STEM images. Similar EELS analysis on an area of MoS$_2$ not in contact with Ti shows no differences between the first and fifth MoS$_2$ layers, as can be seen in Figure S5 in the SI. Thus, only areas of the MoS$_2$ channel directly in contact with Ti show changes in their electronic band structure.

Although the atomic-resolution ADF-STEM images of the interface show clear structural degradations in the first layer, changes in the electronic structure can potentially propagate to deeper layers.

Figure 2(b) shows a set of core-level EELS measurements from each of the first seven $MoS_2$ layers and at one additional position in the Ti contact directly above the $MoS_2$. Changes in the prominence of peaks I and II of the S $L_{23}$ edge were systematically quantified for all the layers (*25*). The fraction of the spectrum at each layer with interfacial character (reference layer 1) is plotted in Figure 2(c). The results show layer 2 and even layer 3 possess considerable interfacial character in their S $L_{23}$ edge. These measurements show the effects of the Ti contact on the $MoS_2$ channel go beyond the surface layer and, consequently, the pinning of the Fermi level likely extends deeper into the layers of $MoS_2$ and is not limited only to the interface as previously speculated (*10, 11*). The effects of STEM probe broadening in the analysis of these results are inconsequential, as the evaluated beam broadening for these experiments was less than 5.5 Å (or less than a $MoS_2$ layer) and affects in quadrature.

Since the crystal structure of the topmost layer of $MoS_2$ is degraded in areas in contact with Ti, the presence of Ti atoms in deeper $MoS_2$ layers cannot be ruled out. Core-level EELS of the Ti $L_{23}$ edge was measured across the first seven $MoS_2$ layers. An example of EELS Ti $L_{23}$ edge dataset can be found in the SI (Figure S4). Figure 2(d) shows the integrated intensity of the Ti $L_{23}$ edge at each layer using the spectrum from layer 0, which is the position directly above the $MoS_2$, as a reference. Surprisingly, in addition to layer 1, layers 2 and 3 also show an appreciable amount of Ti present. The amount of Ti in layers 2 and 3 is significant enough to affect the electronic structure of $MoS_2$ and explain the observed changes in the fine structure of the S $L_{23}$ edge. However, they are not high enough to affect the pristine-like view of the atomic structure as imaged in projection using ADF-STEM. The penetration of Ti into $MoS_2$ can result in reduction of the density of atoms in the Ti sublayer just above the interface, which, as alluded earlier, may be the cause of the reduced contrast observed in ADF-STEM images (Figure 1(d)).

DFT calculations were carried out to determine the structure, bonding and interactions of Ti atoms with $MoS_2$ layers. These simulations examined the systematic addition of individual Ti atoms onto the surface of a monolayer of $MoS_2$ in order to mimic the experimental deposition (see methods and SI for procedure). This "single-atom-addition" approach, while computationally demanding, provides more insight into atomic processes occurring at the metal-$MoS_2$ interface as compared to the addition of whole layers of Ti in a "metal-$MoS_2$-slab" approach (*20-22, 26*).

Figure 3(a) shows the simulated lowest energy structures of the Ti-$MoS_2$ interfaces after the addition one, two, three and five Ti atoms to the top of a $MoS_2$ surface. While many more Ti atoms are involved at the actual metal contact with $MoS_2$, these smaller model systems can provide sufficient insight as to what may be occurring during Ti deposition. The simulations show that the strong interactions between the Ti and S disrupt the $MoS_2$ interface by pulling S atoms out of the $MoS_2$ surface when two or more Ti atoms are present to form a titanium sulfide cluster. These $Ti_xS_y$ clusters are highly stable structures due to the strong Ti-S bond energies. While such interactions lead to a stable Ti-$MoS_2$ configuration, they also destroy the pristine crystal structure of $MoS_2$. This not only disrupts the periodic structure of pure $MoS_2$ but also introduces significant non-periodicity in the Ti clusters above the $MoS_2$ layer. Furthermore, it appears that relatively large openings in the $MoS_2$ layer can also form near the site of the lattice disruption (Figure 3(b)). These "nanopores" in the $MoS_2$ can be large enough to allow Ti atoms to penetrate into deeper layers, which would explain the presence of appreciable amount of Ti in the $2^{nd}$ and $3^{rd}$ layers of $MoS_2$ observed in the EELS measurements. It is worth noting that these simulations were performed without the effects of temperature, which can further enhance destructiveness of the Ti-$MoS_2$ bonds. Ti atoms were added to the system without kinetic energy to intentionally avoid engineering aspects of metal contact deposition. The simulation results demonstrate that the disruption of the $MoS_2$ lattice from the Ti contact is due to the inherent bonding and chemistry of the Ti-$MoS_2$ system and not a result of kinetic energy transfer during deposition as was speculated in literature (*11, 27*).

These DFT calculations also provide a direct comparison between the electronic structures and DOS of a pristine monolayer of $MoS_2$ and that of a five Ti atom-$MoS_2$ interfacial system (Figure 3(c)). Not surprisingly, the Ti interactions with $MoS_2$ significantly alters the electronic structure of $MoS_2$ because of the bonded Ti atoms and loss of symmetry. The total as well as S $3s+3d$ partial DOS of the Ti distorted-$MoS_2$ show a dramatic reduction and broadening of sharp features compared to those of the pristine $MoS_2$, which is consistent with the broadening of peaks I and II in the S $L_{23}$ edge fine structure observed earlier (Figure 2(a)).

The DFT calculations of the metal-$MoS_2$ interface, using the same "single-atom-addition" approach, were extended to metals with significantly weaker metal-sulfur bonds by using Au and In. Both of these metals have significantly lower Gibbs free energies for metal-sulfide formation than Mo or Ti and, therefore, are expected to form very weak bonds with S, which limits the restructuring of the metal-$MoS_2$ interface (*28*). The simulated structures for five Au atom-$MoS_2$ interfaces and five In atom-$MoS_2$ interfaces (Figure 3(d)) show that Au as well as In weakly bind to the $MoS_2$ via weak vdW interactions, which leaves the structure of the $MoS_2$ surface almost unaltered. At first glance, this may appear to be a major advantage having a metal-$MoS_2$ interface with minimum or non-existent Fermi level pinning. However, in this case, the vdW gap acts a tunnel barrier that limits charge transport, which is a fundamental limitation of another kind.

Examining the binding energies of single metal atoms onto the $MoS_2$ surface provides initial insights into metal-sulfur cluster formation and propensity for disruption of the $MoS_2$ surface. The calculations of these binding energies as well as the metal cluster binding energies on $MoS_2$ presented above may provide a way to more rapidly screen and design different metals as well as metal alloys as contact materials with $MoS_2$ that avoid pinning and vdW gaps. Calculated binding energies for a single metal atom on $MoS_2$ for a wide range of different metals are presented in SI (Table S1). Even with just a single-atom, systematic patterns begin to show the similarity of weak binding for In and Au compared to the destructiveness of Ti on $MoS_2$, which is consistent with experimental results. The simulations suggest that metals or metal alloys with more moderate binding energies to $MoS_2$ (those that lie between In (-1.49 eV) and Ti (-3.09 eV)) may be optimal as they would provide a strong enough contact with the $MoS_2$ surface to enable charge injection yet avoid strong covalent bonds that lead to a disruption of the metal-$MoS_2$ interface.

In conclusion, atomic-resolution characterization of a the Ti-MoS$_2$ interface in a FET using ADF-STEM imaging and layer-by-layer EELS shows that the commonly believed premise that improved metal deposition and FET fabrication will solve the pinning problem at the metal-MoS$_2$ interface will need to be reconsidered. Fermi level pinning is likely unavoidable when a metal with high affinity to form strong bonds with sulfur such as Ti is used regardless of deposition engineering. A metal or engineered metallic alloy with just the right affinity to form a moderate chemical bond with sulfur may be a solution to avoid pinning. A simpler way to evaluate the suitability of a metal or an alloy as a contact material with MoS$_2$ that has a potential to avoid pinning or forming van-der-Waals gap, is by calculating the binding energies and atomic structures of a-few-atom big metallic clusters that are brought together on the surface of MoS$_2$.

**Acknowledgements**: STEM analysis was performed in the Characterization Facility of the University of Minnesota, which receives partial support from the NSF through the MRSEC program. We also thank the Minnesota Supercomputing Institute at the University of Minnesota for computational support. This project was partially supported by the MRSEC program of the National Science Foundation under Award DMR-1420013, the C-SPIN, one of the six SRC STARnet Centers, sponsored by MARCO and DARPA and DTRA through Grant No. HDTRA 1-14-1-0042. R.J.W and K.A.M. conceived the study and wrote the paper with assistance and input from S.U., M.N., M.C and Y.W.

**Contributions**: R.J.W. and K.A.M. conceived the study and wrote the paper with assistance and input from S.U., M.N., M.C. and Y.W.  R.J.W. conducted STEM experiments and processed the results with guidance from K.A.M.  S.U. performed all DFT simulations with guidance from M.N. Ti-MoS$_2$ devices were fabricated by R.M. with guidance from S.J.K.

**Supplementary Materials:**

Materials and Methods

Figures S1-S8

Tables S1-S17

Discussion on extension to other metals

References (*29-54*)

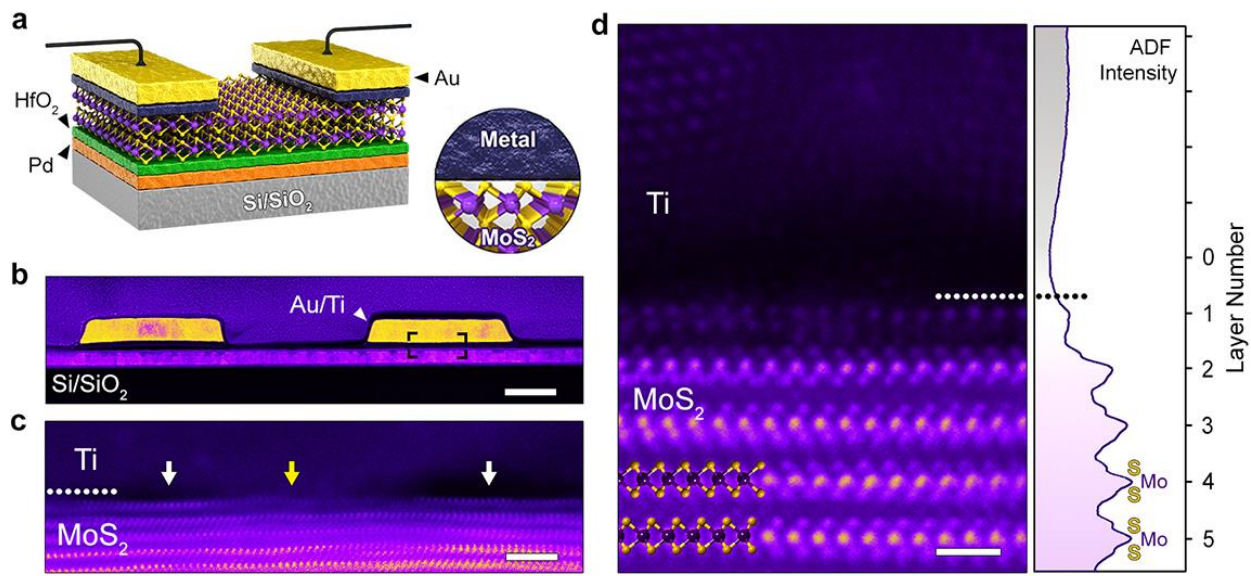

**Fig. 1: Schematic layout and ADF-STEM images of a FET with MoS$_2$ channel: a,** schematic layout of the FET showing the MoS$_2$ channel in contact with Au/Ti metal contacts and the remainder of the device. **b**, low-magnification cross-sectional ADF-STEM image of an actual FET prepared by FIB cutting. The protective amorphous C/Pt layers, also visible here, were deposited on top of the device during TEM sample preparation and are not part of the original FET. Scale bar is 0.2 µm. **c**, high-magnification image of the Ti-MoS$_2$ interface acquired from the boxed area in (b). An area where Ti is clustered is indicated by a yellow arrow, and areas with void pockets are indicated by white arrows. Slight distortions in the lower MoS$_2$ layers are due to small wrinkling of the MoS$_2$. Scale bar is 2 nm. **d**, atomic-resolution ADF-STEM image of the Ti-MoS$_2$ interface acquired from a Ti clustered area. The topmost layer of MoS$_2$ is visibly altered compared to the pristine-like layers below. The horizontally-averaged ADF intensity of the image is shown on the right. A ball-and-stick model of MoS$_2$ is overlayed on the image for clarity of atomic positions. Scale bar is 1 nm.

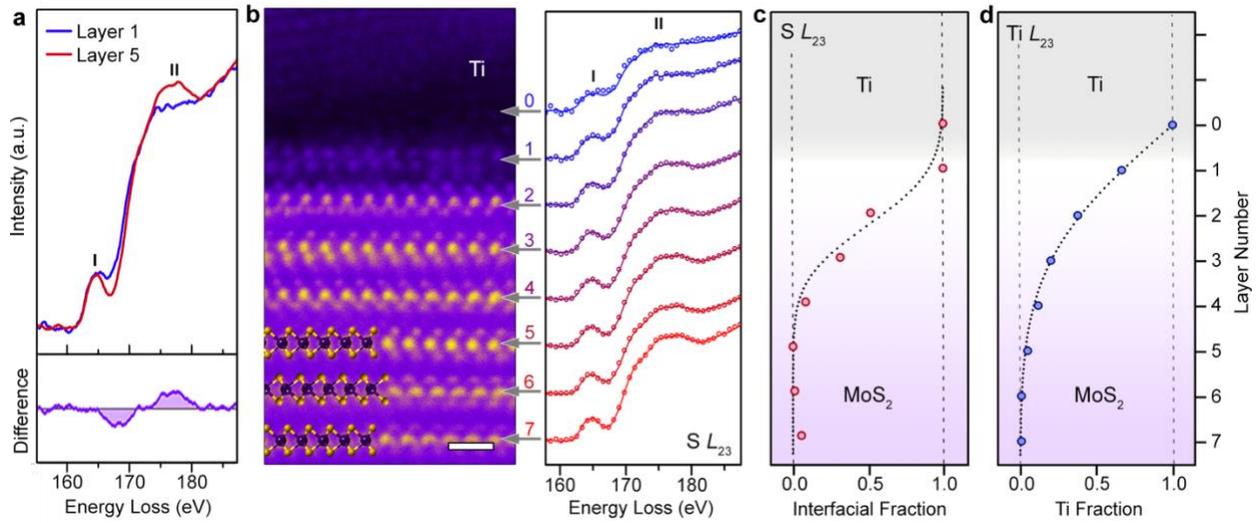

**Fig. 2: Layer-by-layer core-level EELS across the Ti-MoS$_2$ interface: a**, core-level EELS S $L_{23}$-edges measured from the first (nearest to Ti) and the fifth MoS$_2$ layers directly below the Ti contact. The differences between the two spectra are shown below. **b**, (left) atomic-resolution ADF-STEM image of the MoS$_2$ layers and (right) EELS S $L_{2,3}$ edge measured at the corresponding layers shown in the left panel. A ball-and-stick model of MoS$_2$ layers is overlayed on the image for clarity. Scale bar is 1 nm. Measured spectra are shown as scatter points and fitted spectra are shown as lines. **c**, the fractions of the interfacial (reference layer 1) character in their S $L_{23}$ edge in each spectrum obtained from a linear superposition fit of the two reference spectra shown in (a). **d**, the fraction of Ti in MoS$_2$ layers evaluated by integrating the intensity of the measured EELS Ti $L_{23}$-edge obtained at each layer and normalized using the reference spectrum recorded from layer 0. A standard $y = erf(x)$ fit through the data points is shown for (c) and (d).

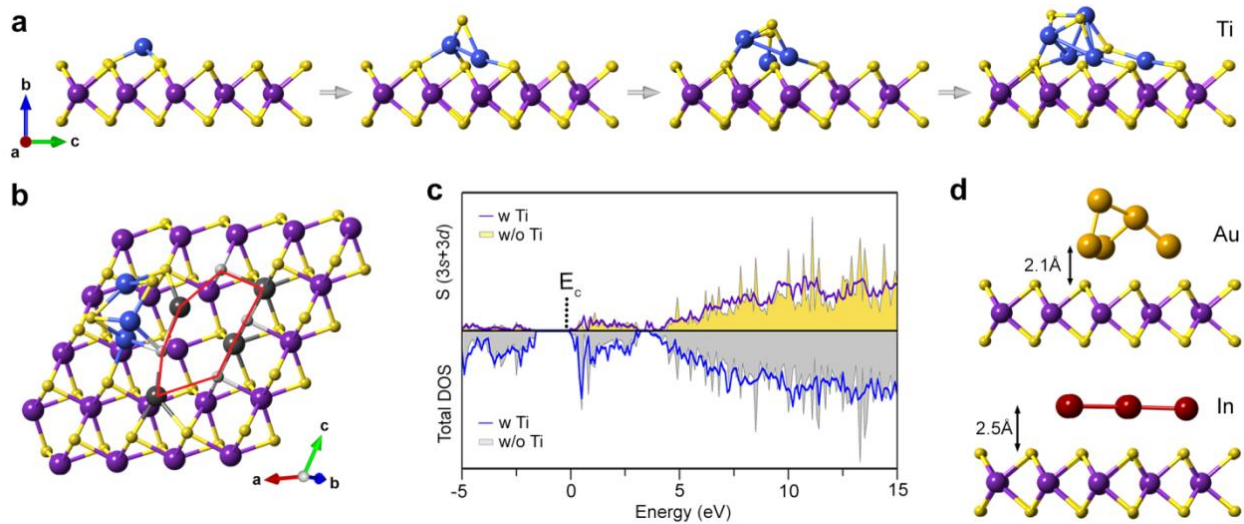

**Fig. 3: Effects of small cluster of metals on atomic and electronic structure of MoS$_2$: a**, ball-and-stick models showing the atomic positions for the lowest energy optimized structures of the Ti-MoS$_2$ system following additions of single Ti atoms up to five Ti atoms as determined by DFT calculations. **b**, a different perspective of Ti-MoS$_2$ system with 5 Ti atoms showing the distortions to the MoS$_2$ lattice. An opening in the lattice is outlined by the grey colored atoms and red line. **c**, the total and S $3s + 3d$ partial electronic DOS of pristine and five Ti atom-distorted MoS$_2$. Sulfur 3$s$ and 3$d$ partial-DOS are the states probed by EELS when measuring S L$_{23}$ edge. The bottom of the conduction band is used as the energy reference point in all cases. **d,** models showing the atomic positions for the lowest energy optimized structures of an Au-MoS$_2$ (top) and In-MoS$_2$ (bottom) system as determined by DFT calculations, similar to (a). In each case, 5 metal atoms and a monolayer of MoS$_2$ were used. The resulting distances between the sulfur atoms of MoS$_2$ and the closest metal atoms of the five-atom clusters are also indicated.